\documentstyle[aps,prl,graphicx,floats]{revtex}

\begin{document}

\wideabs{

\title{Self-similarity of the Third Type in the Strong Explosion Problem}
\author{Andrei Gruzinov}
\address{Center for Cosmology and Particle Physics, Physics Dept., New York University, 4 Washington Place, New York, NY 10003 }

\date{March 10, 2003}
\maketitle

\setcounter{footnote}{0} \setcounter{page}{1}
\setcounter{section}{0} \setcounter{subsection}{0}
\setcounter{subsubsection}{0}

\begin{abstract}

Propagation of a blast wave due to strong explosion in  the center of a power-law-density ($\rho \propto r^{-\alpha }$) spherically symmetric atmosphere is studied. For adiabatic index of 5/3, the solution was known to be self-similar, (of type I) for $\alpha <3$, self-similar (of type II) for $\alpha >3.26$, and unknown in between. We find a self-similar solution for $3<\alpha <3.26$, and give a (tentative) numerical proof that this solution is indeed an asymptotic of the strong explosion. This self-similar solution is neither of type I (dimensional analysis does not work), nor of type II (the index of the solution is known without solving an eigenvalue problem). 
\end{abstract}


}

\textit{Introduction---}
We consider propagation of a blast wave due to strong explosion in  the center of a power-law-density, $\rho \propto r^{-\alpha }$, spherically symmetric atmosphere. This problem is of interest for two reasons. 

An exploding star sends a blast wave into the stellar, and later into the circumstellar medium. The (circum) stellar medium has a varying but spherically symmetrical density, and power-law density profile would often be a reasonable approximation. Blast waves in a circumstellar medium are interesting astronomical sources (eg. \cite{Ch} and later papers by the same author). 

The strong explosion problem is also of methodological interest, being a canonical example of self-similarity in gas dynamics. Self-similar solutions, with shock radius $R\propto t^{\beta }$, of type I and of type II were known in this problem. 

{\bf Type I:} For $\alpha <3$, the blast wave is decelerating, $\beta <1$ \cite{SeB}. The self-similar solution is of type I, meaning that the index $\beta$ is determined by the energy conservation law -- from $R^3\rho (R)(dR/dt)^2 \propto {\rm const}$, one gets $\beta=2/(5-\alpha )$. This is a straightforward generalization of the Sedov-Taylor-von Neumann solution describing the explosion in a uniform atmosphere \cite{Se,Ta,vN,LL}.  

{\bf Type II:} For $\alpha >3.26$ \footnote{Here and in the rest of the paper only the astronomically interesting case of adiabatic index $\gamma =5/3$ is considered. Generalization to other values of $\gamma$ is straightforward.}, the blast wave is accelerating, $\beta >1$ \cite{WS}. The self-similar solution is of type II, meaning that the index $\beta$ is determined by an eigenvalue problem -- it must be chosen so as to remove a singularity of the system of ordinary differential equations describing the self-similar solution. Dimensional analysis is insufficient in this case, and $\beta$ should be determined numerically. 

{\bf Type III:} For $3<\alpha <3.26$, the blast has a constant velocity, $\beta =1$ [this paper]. The self-similar solution is of a new type, type III, meaning that the index $\beta$ is determined by the requirement that the non-self-similar part of the blast wave serve as a piston for the self-similar part of the blast wave. Dimensional analysis does not work in this case, but one can calculate $\beta$ without solving an eigenvalue problem, it is just $\beta =1$.

We consider only a spherically symmetrical model.  In reality, deviations from spherical symmetry should exist \cite{Go}, but are expected to be very small generically (we checked this numerically), should be especially small for  $\alpha \approx 3$, and may be entirely absent for $\alpha > 3$. Gravity is neglected. We first calculate the new self-similar solution, and then describe numerical simulations which confirm that our self-similar solution is indeed an asymptotic of the strong explosion.

\textit{Formulation of the problem}

At $t=0$ and $r\rightarrow \infty$, the atmosphere is spherically symmetrical, cold, motionless, and self-similar with  $\rho \rightarrow r^{-\alpha }$. Also at $t=0$ but at small $r$ there is a non-zero energy density in arbitrary form. The problem is to calculate the resulting flow at $t\rightarrow \infty$. 

\textit{Self-similar solution}

One assumes that at large $t$, the flow is a self-similar blast wave (meaning a shock wave followed by a flow). In dimensionless units, without loss of generality, we will assume that the blast wave propagates into the density
\begin{equation}
\rho_0(r)=r^{-\alpha }.
\end{equation}
The shock radius is
\begin{equation}
R=t^{\beta }.
\end{equation}
The shock jump conditions give the density, velocity, and pressure at $r=R-0$: $\rho = 4\rho _0(R)$, $v=(3/4)\dot{R}$, $p=(3/4)\rho _0(R)\dot{R}^2$. Then the density, velocity, and pressure of the downstream gas are of the form
\begin{equation}
\rho=4t^{-\alpha \beta}\rho (\xi ),
\end{equation}
\begin{equation}
v=(3/4)\beta t^{\beta -1}v (\xi ),
\end{equation}
\begin{equation}
p=(3/4)\beta ^2t^{2\beta -2-\alpha \beta}p (\xi ).
\end{equation}
where $\rho (1)=v(1)=p(1)=1$. 

When (3)-(5) are substituted into the gas dynamics equations (continuity, momentum, and pressure equations), a system of ordinary differential equations for $\rho (\xi )$, $v (\xi )$, and  $p (\xi )$ is obtained. With $'\equiv d/d\xi $,
\begin{equation}
(4\xi -3v)\rho '/\rho =3(v'+2v/\xi )-4\alpha,
\end{equation}
\begin{equation}
(4\xi -3v)v'=p'/\rho+4(1-1/\beta )v,
\end{equation}
\begin{equation}
(4\xi -3v)p'/p =5(v'+2v/\xi )+4(2-2/\beta -\alpha ).
\end{equation}
One can introduce new variables, reducing (6)-(8) to a system of two homogeneous equations \cite{SeB}. We put 
\begin{equation}
v (\xi )=V(\xi )\xi,~~~~~ p(\xi)=\rho (\xi)(C(\xi )\xi )^2.
\end{equation}
Here $C$ is proportional to the sound speed, $V(1)=C(1)=1$, and, after re-defining $'\equiv \xi~d/d\xi =d/d\ln \xi $, we get 
\begin{equation}
V'={F(V,C)\over H(V,C)},
\end{equation}
\begin{equation}
C'={G(V,C)\over (4-3V)H(V,C)},
\end{equation}
where 
\begin{eqnarray}
F=-\left( 15C^2-(4-3V)^2+4(1-\beta ^{-1})(4-3V)\right) V
\nonumber\\
+4\left( (\alpha-2(1-\beta ^{-1})\right) C^2,
\end{eqnarray}
\begin{equation}
H(V,C)=5C^2-(4-3V)^2,
\end{equation}
\begin{equation}
G(V,C)=C\left( F+(6V-4\beta ^{-1})H\right) .
\end{equation}
The system (10),(11) is integrated as follows: start at $\ln \xi =0$ with $V=C=1$, and integrate back to $\ln \xi =-\infty $. The integration can terminate at the singular point $H=0$, unless the singularity is removed by having $F=0$ at this point. \footnote{The singularity $4-3V=0$ with $H\neq 0$ does not occur for the values of $\alpha$ that we consider.}

{\bf Type I, \cite{SeB}:} For $\alpha < 3$, energy conservation gives $\beta =2/(5-\alpha)$. The blast wave is decelerating, $\beta <1$. If also $\alpha > 2$, the solution terminates at the singular point $V=4/3$, $C=0$ at a finite value of $\ln \xi = \ln \xi _{\rm min}$, and an evacuated region forms at small $\xi$. Near the singular point, that is for small $x\equiv \ln \xi -\ln \xi _{\rm min}$, equations (10), (11) give
\begin{equation}
V=4/3-(4/5)(\alpha -1)x, ~~~C\propto x^{1/2}.
\end{equation}
Then (6) gives 
\begin{equation}
\rho \propto x^{-\nu }, ~~~\nu ={8\alpha -18\over 3\alpha -3}.
\end{equation}
For $\alpha <3$, the density slope $\nu <1$. The density singularity is integrable, and the pressure at the singular point $p\propto \rho C^2$ is vanishing. 

{\bf Type II, \cite{WS}:} For $\alpha > 3.26$, $\beta $ is determined numerically, by removing the singularity at $H=0$. The self-similar solution extends all the way to $\ln \xi =-\infty $. The blast wave is accelerating, $\beta >1$.

{\bf Type III, the new regime:} For $\alpha > 3$, the type I solution, $\beta =2/(5-\alpha)$ cannot be correct. This solution has infinite pressure at the termination point at finite $\xi$. Infinite pressure is not necessarily unphysical. We are considering just an asymptotic solution at infinite time. For $\alpha >3$ a finite fraction of the mass of the gas is not described by the self-similar solution, because it originates in the non-self-similar part of the unperturbed density (for a discussion see \cite{WS}). In terms of pure self-similar solution, this gas forms an infinitely heavy piston, which might be able to support the infinite pressure at the end of the self-similar flow. However the problem is that the piston moves at a constant velocity, and therefore it cannot keep up with the accelerated blast wave. We must search among different values of $\beta $. 

For $\alpha > 3$, because of the constant-velocity piston formed by the non-self-similar gas,  the solution cannot be decelerating, that is  $\beta \geq 1$. It turns out (by numerical integration of (10), (11)) that for $\alpha < 3.26$ and $\beta \geq 1$ the solution terminates at the singular point $V=4/3$, $C=0$ at a finite value of $\ln \xi = \ln \xi _{\rm min}$. Near the singular point, that is for small $x\equiv \ln \xi -\ln \xi _{\rm min}$, equations (10), (11) give for $\beta > 1$
\begin{equation}
V=4/3-(8/5)(2-\beta ^{-1})x, ~~~C\propto x^{1/2}.
\end{equation}
Then (6) gives 
\begin{equation}
\rho \propto x^{-\nu }, ~~~\nu =1+{5(\alpha -3)\over 6(2-\beta ^{-1})}.
\end{equation}
For $\alpha >3$, the density slope $\nu >1$. The pressure at the singular point $p\propto \rho C^2$ is diverging. But this is impossible for an accelerating solution, as discussed above. 

We are left with just one possibility, $\beta =1$. In this case, (10), (11) give near the singular point
\begin{equation}
V=4/3-4(1-\alpha /5)x,
\end{equation}
\begin{equation}
C\propto x^{\mu }, ~~~\mu ={\alpha \over 3(5-\alpha )}.
\end{equation}
Then (6) gives 
\begin{equation}
\rho \propto x^{-2\mu }.
\end{equation}
The pressure at the singular point $p\propto \rho C^2$ is finite. Since the blast wave moves at a constant speed, this end pressure can be provided by the piston of non-self-similar gas. Since $2\mu >1$, the mass is diverging, but this is OK, because in terms of an ideal self-similar solution the total mass of the exploding gas is infinite. Part of this mass forms an infinite-mass piston, another part forms an infinite-mass self-similar flow.

The new solution exists only in a narrow interval of density slopes, $3<\alpha <3.26$. For $\alpha >3.26$, numerical integration of (10), (11) with $\beta =1$ does not terminate at the point $V=4/3$, $C=0$. The  solution crosses the line $H=0$ in a different point, with $F\neq 0$. Therefore, our solution is no longer valid. The right solution is the type II solution of \cite{WS}.

\begin{figure}
  \begin{center}
    \includegraphics[angle=0, width=.4\textwidth]{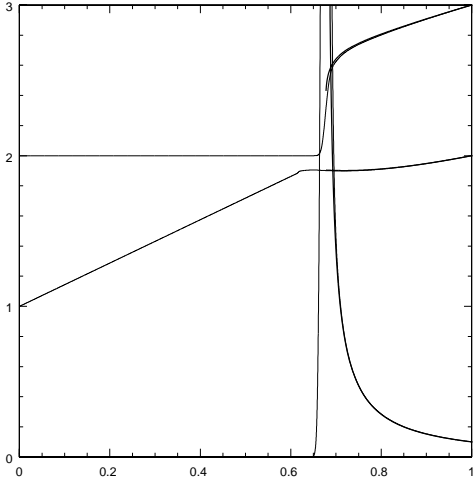}
    \caption{The $\alpha =2.85$ density case. Thick lines show the self-similar solution with $\beta = 2/(5-\alpha )=0.930$. Thin lines show the numerical solution. The horizontal axis gives $r/R_{\rm shock}$. Shown are $0.1\rho /\rho_{\rm shock}$, $1+v/v_{\rm max}$,$2+p/p_{\rm max}$. Resolution and run time: 10000 grid, 20 doublings of the grid. See main text for details.}
  \end{center}
\end{figure}

\begin{figure}
  \begin{center}
    \includegraphics[angle=0, width=.4\textwidth]{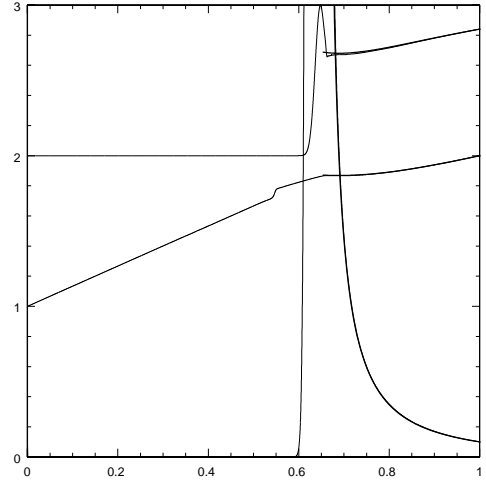}
    \caption{Same as fig. 1, for $\alpha =3.15$, $\beta = 1$. }
  \end{center}
\end{figure}
\textit{Numerical simulation---}
There is no theorem guaranteeing that a self-similar solution is asymptotically reached starting from a generic initial condition. One has to assume this, or  check numerically. We numerically simulated spherically symmetrical explosions in a power-law atmosphere. A spherically symmetric version of the code \cite{TP} was used. It was found that the second-order code becomes unstable near the termination part of the $\alpha \approx 3$ blast waves. We therefore had to use the first-order version of the code. 

It was found that for $\alpha \approx 3$ a very long run is needed to reach the theoretically predicted self-similar solution. This can be explained as follows. The energy conservation gives, approximately, $M(R)\dot{R} ^2=const$, where $M(R)$ is the enclosed mass
\begin{equation}
M(R)=\int_1 ^Rdrr^{-\alpha}\propto R^{3-\alpha}-1.
\end{equation}
Here 1 is the length of the numerical grid, and all the energy of the explosion is put into the first grid point at the initial time. If $\alpha$ is close to 3, one has to wait for a long time, when the shock moves to large $R$, before one of the terms in (22) becomes much larger than the other. Since our new solution exists only in narrow interval of $\alpha$, it is not expected to be very different from the Sedov-Taylor or Waxman-Shvarts solutions. To detect the small difference, we have to  aim at a few percent accuracy for $|\alpha -3|\sim  0.15$. Putting $R^{|3-\alpha |}=30$ gives the greed of $\sim 10^{10}$ points. This is more than we can simulate directly. We used much smaller grids, resolutions of $10^4$ and $10^5$, but then the grid was doubled for up to twenty times, when the blast wave was reaching the end of the grid. 

We first made sure that this numerical method does reproduce the hollow Sedov-Taylor solutions ($2<\alpha <3$, $\beta =2/(5-\alpha)$). It does, as figure 1 shows. We also checked that the numerical method reproduces the Waxman-Shvarts solution (it does for $\alpha =3.40$ with 20 doublings of the grid).

We then checked if the new type III solution ($3<\alpha <3.26$, $\beta =1$) is the true asymptotic of the explosion. It apparently is, as figure 2 shows. However, we have to say that our numerical proof is only tentative. It turns out that increased resolution makes the rear zone of the blast wave (numerically ?) unstable, fig. 3. The density and velocity profiles do not change with increased resolution and still agree with the predicted self-similar profiles. The fluctuating pressure is still centered around the predicted self-similar solution. But the lack of numerical convergence near the end of the blast wave (or a possible indication of a true weak instability?) must somewhat reduce our confidence that our self-similar solution is the true long-time asymptotic of the explosion.

\textit{Conclusion}

A new, type III,  self-similar solution of the strong explosion problem in the power-law atmosphere is given. This solution describes a constant speed ($\beta =1$) blast wave, and applies in a narrow interval of density slopes, $3<\alpha <3.26$. Numerical simulations confirm (tentatively) that the new solution is indeed a long-time asymptotic of the strong explosion. 

\textit{Acknowledgment}

I thank Drs. Goodman, Pen, Sari, Waxman for discussions. This work was supported by the David and Lucille Packard Foundation.

\begin{figure}
  \begin{center}
    \includegraphics[angle=0, width=.4\textwidth]{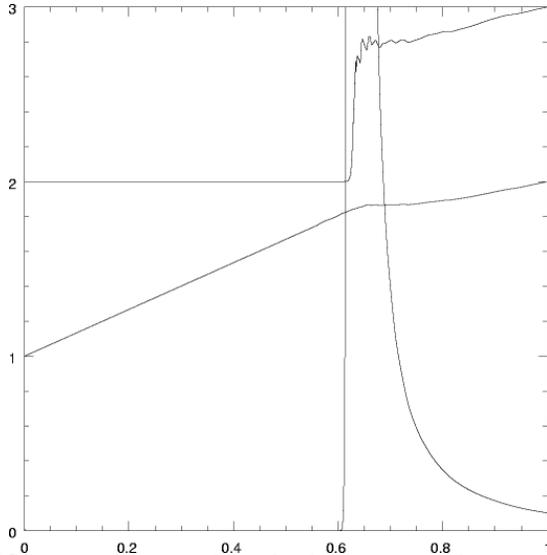}
    \caption{$\alpha =3.15$, analytical solution is not shown, 125000 grid, 18 doublings of the grid.}
  \end{center}
\end{figure}


\begin{thebibliography}{}

\bibitem{Ch}
R. A. Chevalier,
Astrophys. J. {\bf 259}, L85 (1982)

\bibitem{SeB}
L. I. Sedov,
Similarity and Dimensional Methods in Mechanics (New York, Academic, 1959)

\bibitem{Se}
L. I. Sedov,
Prikl. Mat. Mekh. {\bf 10}, 241 (1946)

\bibitem{Ta}
G. I. Taylor,
Proc. R. Soc. London {\bf A201}, 159 (1950)

\bibitem{vN}
J. von Neumann,
Los Alamos Sci. Lab. Tech. Series {\bf 7} (1947)

\bibitem{LL}
L. D. Landau, E. M. Lifshitz,
Fluid Mechanics (Oxford: Pergamon 1959)

\bibitem{WS}
E. Waxman, D. Shvarts,
Phys. Fluids {\bf A5}, 1035 (1993)


\bibitem{Go}
J. Goodman,
Astrophys. J. {\bf 358}, 214 (1990)

\bibitem{TP}
H. Trac, U.-L. Pen,
astro-ph/0210611




\end{thebibliography}
\end{document}